# Strain enhanced electron cooling in a degenerately doped semiconductor

M. J. Prest[1], J. T. Muhonen[1,2], M. Prunnila[3], D. Gunnarsson[3], V. A. Shah[1], J. S. Richardson-Bullock[1], A. Dobbie[1], M. Myronov[1], R. J. H. Morris[1], T. E. Whall[1], E. H. C. Parker[1] and D. R. Leadley[1].

[1]*Department of Physics, University of Warwick, Coventry CV4 7AL, United Kingdom*
[2]*Low Temperature Laboratory, Aalto University, P.O. Box 13500, FI-00076 Aalto, Finland*
[3]*VTT Technical Research Centre of Finland, P.O. Box 1000, FI-02044 VTT Espoo, Finland*

Enhanced electron cooling is demonstrated in a strained-silicon/superconductor tunnel junction refrigerator of volume 40 μm$^3$. The electron temperature is reduced from 300 mK to 174 mK, with the enhancement over an unstrained silicon control (300 mK to 258 mK) being attributed to the smaller electron-phonon coupling in the strained case. Modeling and the resulting predictions of silicon-based cooler performance are presented. Further reductions in the minimum temperature are expected if the junction sub-gap leakage and tunnel resistance can be reduced. However, if only tunnel resistance is reduced, Joule heating is predicted to dominate.

Superconductor tunnel junction refrigerators typically employ a normal-metal, insulator, superconductor (NIS) structure. By biasing the junction close to the superconducting gap it is possible to extract only the most energetic electrons into the superconductor leaving a lower energy, cooled electron gas in the normal metal.[1] Dual junction structures double the cooling power.[2] It is also possible to replace the normal metal with a degenerately doped semiconductor and the insulators with Schottky barriers in a superconductor, semiconductor, superconductor (SSmS) structure. Such a cooler device was demonstrated in Ref. 3 using n+ silicon-on-insulator (SOI) film as the semiconductor and aluminum as the superconductor.

The SSmS structures benefit especially from the low electron-phonon coupling of semiconductors in comparison to metals. However, the SOI based coolers of Ref. 3 suffered from relatively low cooling power, which was attributed to the high tunnel resistances of the Schottky barrier contacts (~ 70 kΩ μm$^2$ for dopant concentration 4x10$^{19}$ cm$^{-3}$) when compared to Cu-Al$_2$O$_3$-Al NIS junctions (to 0.1 kΩ μm$^2$ or below).[4] The Schottky barrier resistance was reduced by increasing the dopant concentration (1.8 kΩ μm$^2$ for a dopant concentration of 1.6x10$^{20}$ cm$^{-3}$), but this approach led to an additional heating mechanism which was most dominant at low bath temperatures.[5] The heating effect was attributed to sub-gap leakage, local phonon heating and back tunneling of quasiparticles from the superconductor.[5,6,7] Similar heating effects have also been attributed to states in the superconductor band gap and to non-equilibrium effects.[4]

Mean-field theory,[8] and our recent electron heating experiments,[9] suggest that strain can dramatically reduce the electron-phonon coupling in silicon due to lifting of the valley degeneracy in the conduction band. Therefore, improved cooling should be possible without having to reduce the tunnel resistance, thereby avoiding heating problems associated with high barrier transparency. In this letter, we report on the cooling performance of such a strained silicon SSmS tunnel junction refrigerator.

The sample preparation is described in detail in our previous publication.[9] The sample layer structures are depicted in Figs. 1(a) and 1(b). The tunnel junction refrigerator is shown in Fig. 1(c). The device has interdigitated contacts with an array of parallel junctions that are used to cool the electron gas in the long thin central mesa. Using an array of small area junctions improves the quasiparticle thermalisation.[10] Also, the total resistance of the silicon is low in this parallel circuit which helps to minimize Joule heating.

The bath temperature $T_b$ of the devices was controlled using a $^3$He/$^4$He dilution refrigerator. The cooler voltage $V_c$ was swept in the range +/- 0.5 mV and the thermometer voltage was monitored whilst biased using a battery powered constant current source (80 pA). The thermometer voltage was calibrated against the RuO thermometer of the dilution refrigerator at zero applied cooler bias, for a range of bath temperatures. At low temperatures the probe voltage tends to saturate[11] and therefore we had to extrapolate the calibration at the lowest temperatures (below ~200 mK).

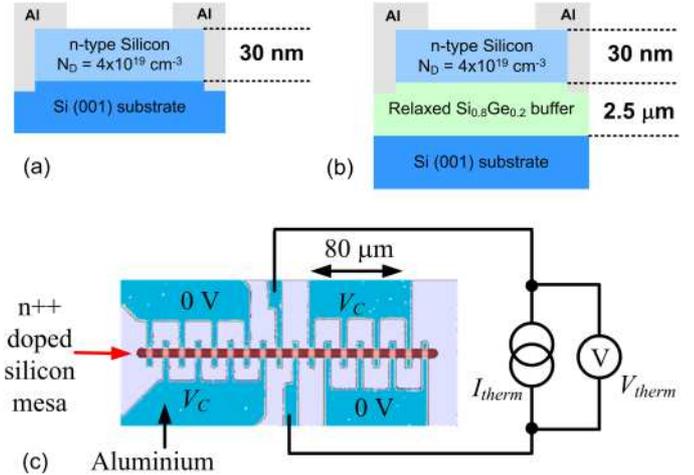

Figure 1. (Color online) a) control sample layer structure b) strained sample layer structure c) Optical image of the cooler device (colors have been altered). The interdigitated aluminum (superconductor, thickness 200 nm) cooler contacts are located at the left and right hand side of the central junction pair, which is used as a thermometer, giving the temperature of the electron gas in the middle of the mesa. The mesa is approximately 100 nm high and the electron gas is restricted to the degenerately doped region within the top 30 nm. The length and width of the mesa are 222 μm and 6 μm, respectively. The junction areas are all 6 μm x 6 μm. The cooler voltage $V_C$ is applied between the pairs of junction arrays and the two sets of arrays are biased so that there is no parasitic current through the thermometer. The thermometer is biased using a high impedance battery powered current source ($I_{therm}$) and its voltage ($V_{therm}$) is monitored using a floating differential voltage amplifier.

Using the calibration we have converted probe voltages to temperatures and plotted the electron temperature $T_e$ versus the applied cooler bias $V_c$ for discrete $T_b$ in Fig. 2. The main part of Fig 2 shows the cooling curves of control and strained sample at ~300 mK. The inset shows the full set of the curves for the strained sample (control not shown). In each case cooling can clearly be seen as the bias approaches +/- 0.36 mV. At the lowest $T_b$ there is an initial increase in $T_e$ for low $V_c$ (inset of Fig. 2). The low temperature heating is thought to be a combination of heat returned by quasi-particles and cold electron tunneling due to the sub-gap leakage path.

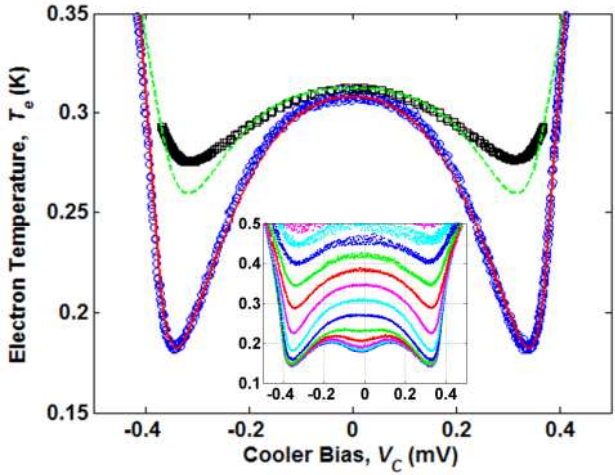

Figure 2. (Color online) Electron temperature $T_e$ versus cooler bias $V_C$. The black squares are experimental data for the unstrained control sample. The green broken curve is a fit to the control data. Blue circles represent experimental data for the strained sample and the red curve is a fit to the model. Fit parameters are given in the text and in Table 1. The inset shows $T_e$ versus $V_C$ at different bath temperatures for the strained sample. Note that $T_b=T_e$ when $V_C = 0$.

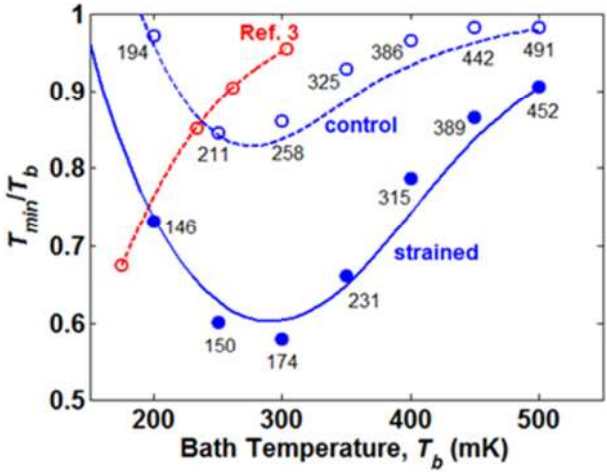

Figure 3. (Color online) Cooling performance of strained (solid circles) and unstrained (open circles) samples for a range of bath temperatures $T_b$. Minimum electron temperature $T_{min}$ at each $T_b$ is obtained from the $T_e$-$V_c$ cooling curves (see Fig. 2). The values shown by the points are $T_{min}$ in mK. Solid blue line: calculated $T_{min}/T_b$ for the strained sample. Broken blue line: calculated $T_{min}/T_b$ for the control sample. Red points from Ref. 3, broken red line is a guide to the eye. Junction cooling powers $P_C$ were calculated for $V_C = 0.323$ mV.

The ratio $T_{min}/T_b$, where $T_{min}$ is the minimum temperature on each $T_e$-$V_c$ cooling curve, can be considered as one figure of merit of a cooler. This ratio is plotted as a function of $T_b$ in Fig. 3. The strained electron cooler has an optimum cooling performance at a bath temperature of 300 mK, reducing the electron temperature to 174 mK. Earlier SOI coolers[3] with the same dopant concentration did not cool well from 300 mK, but cooled better at the lowest temperatures, see Fig. 3. The more effective cooling of the SOI based device at lower temperatures is attributed to its lower sub-gap leakage. The SOI coolers had a similar volume (5% larger) and samples with a higher dopant concentration[5] of $6.7 \times 10^{19}$ cm$^{-3}$ (and smaller volume by a factor of 0.87) showed similar cooling to our control sample with an optimum $T_{min}/T_b$ of about 0.77 at 250 mK.

We have modeled the cooling power $P_c$ and current $I$ in our devices using equations,[7, 12]

$$P_c = \frac{2}{e^2 R_T} \int_{-\infty}^{\infty} (E - eV_C/2)$$
$$f(E - eV_C/2, T_e) - f(E, T_b) \, g(E, \Gamma) \, dE \quad (1)$$

$$I = \frac{1}{e R_T} \int_{-\infty}^{\infty} f(E - eV_C/2, T_e) - f(E, T_b) \, g(E, \Gamma) \, dE \quad (2)$$

where $f(E, T)$ is the Fermi distribution function and

$$g(E, \Gamma) = \left| \text{Re} \left[ \frac{E + i\Gamma}{\sqrt{(E + i\Gamma)^2 - \Delta^2}} \right] \right|$$

is the Dynes density of states.[13] We assume that the Fermi distributions in the superconductor and semiconductor are set by $T_b$ and $T_e$, respectively. $R_T$ is the junction tunnel resistance, i.e. half of the total (or normal state) resistance of the cooler junctions. $V_C$ is the total voltage applied across the cooler junctions. The $\Gamma$ parameter represents the broadening of the density of states and emergence of states in the energy gap of the superconductor. This parameter is often used as a figure of merit for NIS type junctions ($\Gamma/\Delta$ defines the magnitude of the sub-gap leakage). $\Gamma$ has been found to be useful parameter when modeling the characteristics of NIS junctions and in this context it does not necessarily represent a change in the nature of the superconductor density of states.[14] As mentioned earlier, similar characteristics have been modeled as non-equilibrium or quasiparticle related effects.[4, 7]

Electron cooling is opposed by heat flow from the lattice phonons $P_{e-ph}$. In single (strained) and many-valley (unstrained) semiconductors $P_{e-ph}$ can be approximated by a power law[13]

$$P_{e-ph} = \Sigma v (T_e^n - T_b^n), \quad (3)$$

where pre-factor $\Sigma$ and power $n$ depend on the microscopic parameters and $v$ is the volume of the electron system. It has been found that $n = 6$ can be used for strained and unstrained silicon structures.[9] We have determined $P_{e-ph}$ by integrating the electron-phonon conductance, ($G_{e-ph} = dP/dT$) as measured in our previous work,[9] then fitting $T^6$ power laws to $P_{e-ph}$ of the control and strained samples, respectively.

Other mechanisms considered which oppose the cooling are quasiparticle back-tunneling and recombination[15] $P_{QP}$, and Joule heating in the semiconductor $P_J$

$$P_{QP} = -\beta(IV + P_{cool}) \quad (4)$$

$$P_J = -I^2 R_{Sm} \quad (5)$$

Here $\beta$ is the fraction of power returned by quasiparticle back-tunneling and recombination and $R_{sm}$ is the semiconductor resistance. The electron temperature is then obtained from a heat balance equation

$$P_c + P_{e-ph} + P_{QP} + P_J = 0 \quad (6)$$

The strained and control cooler data were fitted to the above model using the parameters given in Table 1, as shown in Figs. 2 and 3. At a $T_b$ of 300 mK, the strained sample has an excellent fit (Fig. 2) but the control fit is not so good. The control sample has a lower tunnel resistance and therefore causes higher power dissipation in the superconductor. This leads to quasiparticle heating of the superconductor which is not well accounted for in the current model. However, Fig. 3 shows that the fits at other bath temperatures are reasonable for both samples and the model seems to capture the characteristic behavior of strained and unstrained samples. For

Fig. 3 the $T_{min}$ values were calculated at bias $V_C = 0.85 \times 2\Delta/e = 0.323$ mV, which is close to the optimum bias.

Figure 4 shows calculated junction cooling and opposing heating powers for strained and control devices at a $T_b$ of 300 mK using values from table 1. $P_C$ (black, 1) is much smaller for the strained sample as a result of its higher tunnel resistance. The lower $P_{e-ph}$ (green, A) of the strained sample, however, means that less cooling power is required to cool the electrons. The minimum electron temperatures are found at the intersection of the heating and cooling powers, equivalent to solving the heat balance equation. We have highlighted the $T_{min}$ values for the full heat balance expression, as given in equation 6. The red (B) and blue (C) lines show the effect of the increased heating due to Joule heating and Joule heating plus quasiparticle heat return, respectively. Note that Joule heating is a stronger heating mechanism in the control sample, which is attributed to higher currents as a result of the lower $R_T$. The magenta curves (2) show $P_C$ for $\Gamma = 0$, to illustrate the ultimate cooling limit for these $R_T$ values.

Table 1. Model parameters used for Fig. 2. Also $\Delta = 0.19$ meV, $\nu = 4 \times 10^{-17}$ m$^3$ and $\beta = 0.02$ for both devices. For the strained (control) sample the sheet resistance was 569 (383) $\Omega$/square.[8] $R_{sm}$ results from 14 squares in parallel between the cooler contacts.

|  | $R_T$ $\Omega$ | $R_{sm}$ $\Omega$ | $\Sigma$ WK$^{-6}$ m$^{-3}$ | $\Gamma$ $\times \Delta$ |
|---|---|---|---|---|
| Strained | 400 | 41 | $2 \times 10^7$ | 0.01 |
| Control | 40 | 27 | $5.2 \times 10^8$ | 0.015 |

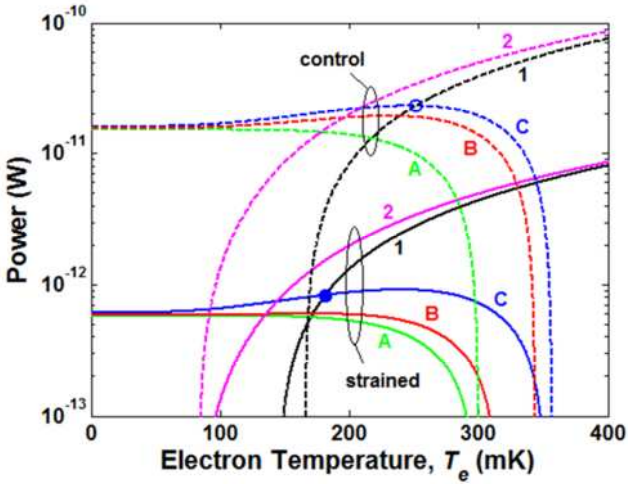

Figure 4. (Color online) Calculated cooling power versus electron temperature $T_e$ for control (broken curves) and strained (solid curves) devices at a bath temperature $T_b$ of 300 mK. Black curves (1): Junction cooling power $P_C$ at $V_C = 0.323$ mV using $\Gamma$ from table 1. Magenta curves (2): Junction cooling power $P_C$ at $V_C = 0.323$ mV for $\Gamma = 0$. Green curves (A): $P_{e-ph}$. Red curves (B): $P_{e-ph}+P_J$. Blue curves (C): $P_{e-ph}+P_J+P_{QP}$. Filled circle: $T_{min}$ for the strained device. Open circle: $T_{min}$ for the control device.

We have investigated the optimal cooler junction resistance by calculating the minimum electron temperature expected in the strained sample for various tunnel resistances $R_T$, using $\Sigma = 2 \times 10^7$ WK$^{-6}$m$^{-3}$. Figure 5 shows the influence of the various heating mechanisms. If we assume an ideal junction ($\Gamma = 0$) and only heating from the lattice (red curve, 1) then cooling is improved for the lowest tunnel resistances. If we introduce a non-zero $\Gamma$ (green curve, 2), some additional heating occurs. The addition of Joule heating (blue curve, 3) shows that low tunneling resistances lead to poor cooling performance as the Joule heating becomes dominant. The inclusion of reduced cooling due to quasiparticle heat return (black line, 4) simply increases the temperature further; we assume this mechanism is weak in our samples due to the array of small area junctions. The minimum temperature of our device, with an $R_T$ of 115 k$\Omega\mu$m$^2$, lies just to the right of the lowest predicted minimum temperature of the full model (black curve) and is shown as a cross. Note that curve 1 of Fig. 5 suggests that high quality junctions with low sub-gap leakage and $R_T$ less than about 20 k$\Omega\mu$m$^2$ could allow cooling from 300 mK to below 100 mK in a strained Si cooler device.

To summarize, the lifting of the valley degeneracy of the conduction bands in strained silicon causes a decrease in the electron-phonon coupling.[8, 9] This leads to lower electron temperatures in tunnel junction refrigerators because the cooling power is opposed by less heat from the lattice. The cooling power required to reach low temperatures is thus also reduced and helps to compensate for the high tunnel resistances of silicon-superconductor tunnel junctions in comparison to more conventional, metal based NIS junctions. Further improvements in cooling are expected, particularly at the lowest temperatures, if sub-gap leakage ($\Gamma$ values) can be reduced in SSmS tunnel junction refrigerators.

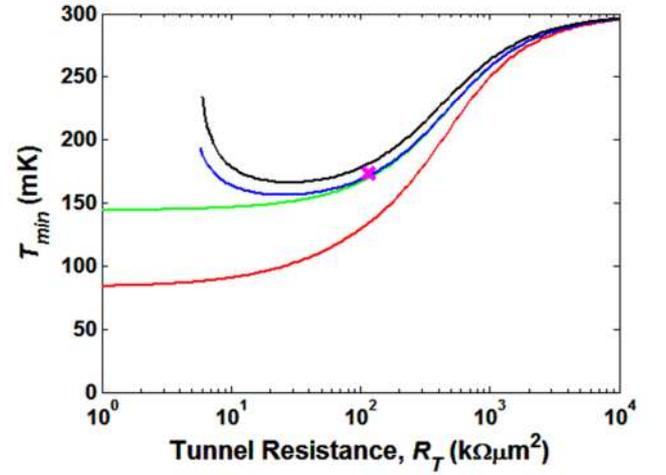

Figure 5. (Color online) Calculated $T_{min}$ versus $R_T$ using different cooling calculations for the strained sample. Red curve (1): $P_C + P_{e-ph} = 0$, $\Gamma = 0$. Green curve (2): $P_C + P_{e-ph} = 0$, $\Gamma = 0.01 \Delta$. Blue curve (3): $P_C + P_{e-ph} + P_J = 0$, $\Gamma = 0.01 \Delta$. Black curve (4): $P_C + P_{e-ph} + P_J + P_{QP} = 0$, $\Gamma = 0.01 \Delta$, $\beta=0.02$. The magenta cross shows the cooling measurement result for the strained sample. Junction cooling powers $P_C$ were calculated for $V_C = 0.323$ mV.

We gratefully acknowledge J. P. Pekola, M. Meschke and A.M. Savin for fruitful discussions. This work has been financially supported by EPSRC through Grant No. EP/F040784/1 and by EC through Projects 228464 MICROKELVIN and 257375 Nanofunction Network of Excellence and by the Academy of Finland.